# Mechanical resonance: 300 years from discovery to the full understanding of its importance

Jörn Bleck-Neuhaus

Department 1 of Physics and Electrical Engineering

University of Bremen, PO box 330440

D-28334 Bremen, Germany

Email: bleck@physik.uni-bremen.de

Orcid-ID: 0000-0002-3226-8357

**Nov 18th , 2018**

Abstract

Starting from the observation that the simplest form of forced mechanical oscillation serves as a standard model for analyzing a broad variety of resonance processes in many fields of physics and engineering, the remarkably slow development leading to this insight is reviewed. Forced oscillations and mechanical resonance were already described by Galileo early in the 17[th] century, even though he misunderstood them. The phenomenon was then completely ignored by Newton but was partly rediscovered in the 18[th] century, as a purely mathematical surprise, by Euler. Not earlier than in the 19[th] century did Thomas Young give the first correct description. Until then, forced oscillations were not investigated for the purpose of understanding the motion of a pendulum, or of a mass on a spring, or the acoustic resonance, but in the context of the ocean tides. Thus, in the field of pure mechanics Young's results had no echo at all. On the other hand, in the 19[th] century mechanical resonance disasters were observed ever more frequently, e.g. with suspension bridges and steam engines, but were not recognized as such. The equations governing forced mechanical oscillations were then rediscovered in other fields like acoustics and electrodynamics and were later found to play an important role also in quantum mechanics. Only then, in the early 20[th] century, the importance of the one-dimensional mechanical resonance as a fundamental model process was recognized in various fields, at last in engineering mechanics. There may be various reasons for the enormous time span between the introduction of this simple mechanical phenomenon into science and its due scientific appreciation. One of them can be traced back to the frequently made neglect of friction in the governing equation.

# 1 Forced oscillations and resonance today

Forced oscillations with their characteristic resonance phenomenon have widespread significance in physics and engineering. While the term *resonance* literally means audible repercussion and originally referred to the familiar acoustic phenomenon, at the end of the 19$^{th}$ century this concept began to invade other fields of physics and technology. Today it is employed wherever forced oscillations are observed, and with this broad significance it will be used in the present paper. Examples outside acoustics include the following:

(1) In mechanics, forced oscillations and resonance frequently serve as an example in academic textbooks on Newton's laws of classical mechanics, but they also provide an indispensable criterion for the design of high buildings, long bridges, ocean liners swimming stably upright, or rapidly spinning drive axles.

(2) In electrodynamics, forced oscillations and resonance provide the key to understanding wireless communication technology, or magnetic resonance tomography as used in medical diagnostics, or resonance absorption spectroscopy, be it with microwaves for monitoring environmental trace gases or with gamma rays for solid state physics in the Mössbauer effect.

(3) In quantum mechanics, resonance is of fundamental importance in the formation and decay of any unstable physical system.

The use of the same term for such a broad variety of phenomena finds its justification in that, independent of their nature, forced oscillations and resonance of all kinds mentioned share a common ground. They all follow the same equations and therefore can be understood in terms of the simplest possible example. This example consists of a single body, elastically bound to a position of rest and subjected to friction and to a force periodically varying in time. Typically, this system is characterized by its *resonance curve*. To illustrate its widespread significance, Richard Feynman, in his legendary *Lectures on Physics,* coined the bon mot that there was no volume of *The Physical Review*, the biggest journal dedicated to original research in physics then, without at least one figure showing a *resonance curve* (Feynman et al. 1965, ch. 23). He put his assertion to a test with the then latest volume of the journal and found, indeed, two resonance curves (see Fig. 1). In this case, they proved by quantum mechanical reasoning the existence of an unstable elementary particle unknown hitherto.

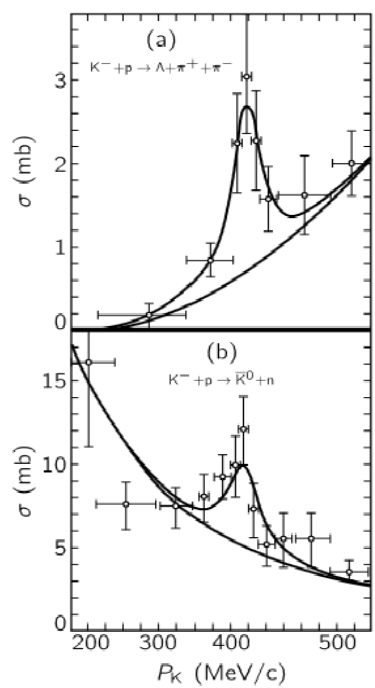

Fig. 1 The resonance curves shown by Feynman (Ferro_Luzzi 1962).

Accordingly, every textbook of physics or engineering, if it has a section on mechanics, also has a chapter dedicated to forced mechanical oscillations (e.g. Meschede 2015; Szabo 2013a). To a typical physicist or engineer, the appreciation of this topic as a fundamental one might appear natural, and certainly as something established a long time ago. However, this was not so.

It is true that forced oscillations have been known in modern science from the very beginning. It seems, however, that in early classical mechanics the question of how oscillations can be excited, other than by a single push, was perceived as quite unimportant. Remarkably, the relevant formulae were discovered long after the advent of classical mechanics. Moreover, these discoveries were not made in connection with simple oscillations of pendulums or springs, as they are presented in textbooks today, but in a far more difficult context, namely, the theory of ocean tides. In shipping activities it is obviously important to know in advance the height and schedule of high and low water at ports and coasts. Accordingly, observational data have been collected since ancient times (Cartwright 2000).

After this early perception, the concept of forced mechanical oscillation and resonance began to attract appropriate interest only after almost 300 years. Then, however, interest grew so rapidly that in 1913 even the Nobel committee considered to award the physics prize for a single skillful application of mechanical resonance. Nevertheless, it still took additional decades for the simple archetype of mechanical resonance of an oscillating single mass on a spring to gain the full attention it receives today.

When judged today, this 300 year development might seem remarkably slow, compared e. g. with the rapid development of celestial mechanics in the the much shorter period of only the 18$^{th}$ century (Laplace 1799). This is much in contrast to the more recent case of another type of resonance, i. e. electromagnetic resonance, whose rapid reception within only a few decades has been documented by Blanchard (Blanchard 1941). No corresponding historical review of mechanical resonance seems to exist. References to studies of mechanical resonance were rare for a long time and mostly in error.

The present article aims to make a first contribution to fill this gap. Owing to the absence of secondary literature, the study had to be based mainly on primary sources. It is seen that even within the field of mechanics the process of understanding the resonance phenomenon was far from straightforward. There were obstacles like misunderstandings, ignorance of correct treatments achieved previously, and several independent (re-)discoveries. The article is based on literature easily accessible and emphasizes German sources from the mid-19[th] century on, supposing that the development was similar in other places. It is hoped that the material presented may help future efforts to conduct a more comprehensive historical analysis.

## 2   Physics aspects

### 2.1   The mechanical system

Since any type of forced oscillation and resonance today is described in the same way as in mechanics, we will first review the physical characteristics of this simple model. The model is the one-dimensional motion of a single body, bound by an ideal spring to a position of rest, subjected to friction and acted on externally by a driving force that varies harmonically (i.e. sinusoidally) in time. The corresponding equation of motion reads:

(eq. 1) $$m\frac{d^2s}{dt^2} = -k\,s - c\frac{ds}{dt} + F_1 \sin \Omega\, t$$

The variables are $t$ - time, and $s$ - displacement from rest position. The parameters of the oscillating system are: $m$ - mass; $k$ - strength of spring; $c$ - strength of friction. Parameters of the driving force are: $F_1$ - maximum strength; $\Omega$ - angular frequency, so that the disturbance by the force has the period $T_{ext.force} = 2\pi/\Omega$. Derived parameters of importance are: $\omega_0 = \sqrt{k/m}$ - angular frequency of free proper oscillation in the absence of forcing and damping; $\beta = c/2m$ - damping constant for the attenuation of the free oscillation; $\omega_d = \sqrt{\omega_0^2 - \beta^2}$ - angular frequency of free damped oscillation; $Q = \omega_0/2\beta$ - dimensionless factor characterizing the effect of damping on resonance and transient oscillations. Typically, we have the case of weak damping, i. e. $Q \gg 1$. Note that the zero of the time scale is definitely chosen so that $t = 0$ is the moment of zero external force. In the following we will drop the adjective *angular* and speak only of *frequency*. The model system is referred to as the *oscillator*.

## 2.2   Forced oscillation

Possible movements of the oscillating body are given by the general solution to the differential equation of motion (eq. 1). It reads, except for one special case (see below eq. 5):

(eq. 2) $$s(t) = A \sin(\Omega t - \delta) + B\, e^{-\beta t} \sin(\omega_d t - \alpha) \; .$$

Both terms on the right represent harmonic oscillations, albeit of quite different significance. The first term is called the *stationary forced oscillation* because it represents the latter in the literal sense. The second term represents a *damped free oscillation*. While the stationary forced oscillation is uniquely defined by the parameters of the system and maintains a nonzero amplitude, the damped free oscillation may be present or be absent ($B = 0$) depending on the particular *initial conditions*, i. e. the position and velocity of the oscillator at a freely chosen time $t = t_0$.

The amplitude of the stationary forced oscillation, usually expressed as function of $\Omega$, is given by

(eq. 3) $$A(\Omega) = A_{static} \, \frac{\omega_0^2}{\sqrt{(\Omega^2 - \omega_0^2)^2 + (2\beta\Omega)^2}} \; ,$$

where

(eq. 3a) $$A_{static} = \frac{F_1}{k}$$

is the displacement of the mass if a steady force $F_1$ is applied.

The phase shift $\delta$, also dependent on $\Omega$, but independent of the amplitude $F_1$, is

(eq. 4) $$\delta(\Omega) = \tan^{-1} \frac{2\beta\Omega}{\omega_0^2 - \Omega^2} \ .$$

The damped free oscillation can modify the motion in different ways, depending on its amplitude $B$ and phase $\alpha$ which are defined so that $s(t)$ matches the particular initial conditions. With properly chosen initial conditions, even the case $B = 0$ is possible, reducing $s(t)$ to the pure stationary forced oscillation. Note that the frequency $\omega_d$ of the damped free oscillation is independent of its amplitude, which is one of the defining characteristics of the harmonic oscillator.

The phenomenon as a whole is characterized by the *resonance curve* (as addressed by Feynman), which is the graph of the amplitude function $A(\Omega)$. The appearance of this function is largely determined by the strength of damping as expressed by the $Q$-factor. In the typical case, damping is weak (i.e. $2\beta \ll \omega_0$, or equivalently $Q \gg 1$), and the resonance curve shows a sharp maximum of width $\Delta\Omega = \omega_0/Q$ in the region where $\Omega \cong \omega_0$. This can easily be experienced, e.g., with a pendulum swinging in air with a $Q$-factor in the range $10^2 - 10^3$. At resonance, the amplitude is $Q$ times bigger than the amplitude $A_{static}$ which would result if the driving force varied slowly (i.e. $\Omega \ll \omega_0$). Correspondingly, the energy content of the forced oscillation at resonance is $Q^2$ times the energy content of the static displacement. Regarding the possibly big values of these enhancement factors it is clear that not every mechanical oscillator can replicate the entire resonance peak. In the worst case the oscillator will be destroyed in a *resonance disaster* when the driving frequency $\Omega$ approaches the resonance condition $\Omega \cong \omega_0$.

The only case of exception to eqs. 2 to 4 is the *exact* resonance of the *undamped* oscillator, mathematically described by letting $\omega_0 = \Omega$ and $\beta = 0$. Then we have

(eq. 5) $$s(t) = A_{static} \frac{\omega_0 t}{2} \sin(\omega_0 t - 90°) + B \sin(\omega_0 t - \alpha)$$

Again, the first term on the right is the forced oscillation and the second term a free oscillation which now is undamped. The forced oscillation is not stationary but has an amplitude rapidly growing in time. After only one period it exceeds the static displacement by a factor of $\pi$ already. No real system can behave in this way for long times. It will eventually pass its limits of elastic response and self-destruct.

## 2.3 Effects of friction

The spectacular resonance catastrophe at driving frequency $\Omega = \omega_0$ (eq. 5) is not the only consequence of the absence of friction. At any driving frequency, the system behaves quite differently when friction is present ($c > 0$) than when it is truly absent ($c = 0$). In the case with friction, every possible motion of the system will pass through a transient state, which strongly depends on the initial conditions, to sooner or later approach the stable stationary harmonic oscillation which is independent of the initial conditions. This is because the superimposed free oscillation fades away with time constant $1/\beta$. The $Q$ factor gives the number of oscillations after which the superimposed free oscillation has dropped to about 4% (exactly: $e^{-\pi}$) of its initial amplitude. Hence, the weakly damped oscillator in its transient state can show motions of more or less complicated shapes which certainly are not intuitively perceived as a direct consequence of the initial conditions. To see that, imagine only some fixed values of position and velocity, but different times $t_0$ at which these values are imposed as initial conditions. (Remember that a moment of zero driving force marks the time $t = 0$.) Notwithstanding its importance, the transient phase is most often excluded in textbooks and elsewhere. Those treatises usually jump directly to the final state, i. e. the unique stationary oscillation.

In a system without friction, on the other hand, the transient phase never ends. This is seen from eq. 2 by letting $\beta = 0$, because then the amplitudes $A$ and $B$ of both oscillations remain constant. Unless the two frequencies $\Omega$ and $\omega_0$ are in proportion of small natural numbers, or one of the two amplitudes is much smaller than the other, this motion can appear quite irregular, depending on the initial conditions, and remains so for all times.

## 2.4 Phase shift $\boldsymbol{\delta}$ and energy transfer

In stationary oscillation, the phase shift $\delta$ controls the energy exchange between the oscillator and the mechanism that produces the driving force. In resonance ($\Omega = \omega_0$) the phase shift is $\delta = 90°$, or a quarter of a period. This holds independently of whether the oscillator is damped or not. It means that the direction of the body's velocity changes sign when the external force does, because the body's velocity also has a 90° phase shift with respect to its position. Consequently, the external force acts permanently in the direction of motion and so does positive work on the oscillator.

For any other driving frequency $\Omega \neq \omega_0$, the absence or presence of damping makes an important difference which for long time was overlooked. In the frictionless case, the phase shift is $\delta = 0°$ for driving frequencies below resonance, and $\delta = 180°$ above. That means that, without damping, at frequency $\Omega < \omega_0$ the oscillator follows the variations of the force without any delay (formerly termed "*direct oscillation*"), while at frequency $\Omega > \omega_0$ the motion is exactly contrary to the direction of the force (formerly "*indirect oscillation*"). In both cases the energy transferred in one quarter of each period is exactly given back in the next quarter. The average work done by the driving force is always zero.

With friction present, however, the phase $\delta$ lies truly inside one of the intervals $0° < \delta° < 90°$ or $90° < \delta° < 180°$. Therefore the external force and the velocity of the body invert their directions not at the same times. Again, in each period the energy flux reverses its direction four times, but now leaves a net energy transfer to the oscillator,

(eq. 6) $$E_{transferred}(\Omega) = \pi\, F_1\, A(\Omega)\, \sin \delta(\Omega)\,.$$

This energy transfer is always positive. It exactly balances the energy loss due to friction, thus rendering the stationary oscillation stationary.

These considerations hold in the stationary state but not during the transient phase. In the transient phase the phase shift between oscillator and driving force can vary from one oscillation to the other, causing the energy transfer to deviate strongly from eq. 6, both in magnitude and direction.

## 2.5  Non-harmonic excitation

It is for mathematical reasons that a driving force $F(t)$ with sinusoidal variation leads to the simplest case of forced oscillations. In practice, however, a given periodic driving force rarely varies sinusoidally. Harmonic analysis then shows that the force has harmonic overtone components, each one having its particular amplitude and phase shift. Every component varies sinusoidally, but at a frequency $n\Omega$ (with natural number $n \geq 2$). As a consequence of the linearity of eq. 1, a unique stationary forced oscillation still exists but now is given by linear superposition of all stationary forced oscillations (first term in eq. 2) that would correspond one by one to the harmonic overtone components of the force. However, since amplitude and phase shift of each of these harmonic oscillations vary with overtone number $n$ (through the frequency $n\Omega$ inserted into eqs. 3 and 4), their superposition can build up to form a motion of a shape which strongly differs from the way the driving force varies in time.

An extreme example for non-harmonic excitation is the familiar case of short pushes repeated at time intervals approximately $n$ times longer than the period of the pendulum. All parents do so when they intermittently push their child sitting on a swing. In this case an overtone of the series of pushes is nearly in resonance with the free oscillation of the pendulum. If the said superposition of oscillations is worked out, the result shows that the pendulum, in the pauses between the pushes, oscillates with its proper frequency $\omega_0$ as if it were free, while after approximately $n$ such oscillations a push causes a slight phase jump. Put strictly, the motion of the swing indeed is periodic, albeit with frequency $\Omega$ instead of $\omega_0$.

## 3   Galileo in error with resonance

The phenomenon of forced oscillations and mechanical resonance is a fairly familiar one and must have been known already to prehistoric people. Imagine how easy it is, when sitting on a tree branch that is swinging up and down, to unintentionally provoke stronger oscillations which may lead to a resonance disaster with potentially harmful consequences.

Following Truesdell (Truesdell 1960 p. 34, 2012 p. 323), forced oscillation and resonance had their earliest clear appearance in the scientific literature when Galileo (1564–1642) discussed in 1638 that a weak force can move an oscillatory system much farther from its rest position when it is applied intermittently rather than constantly (Galilei 1638). Galileo mentions examples like pendulums, church bells, and strings of musical instruments. The last example also spawned the word *resonance*.

To Galileo, this phenomenon appeared as a strange *enhancement of force*, illustrated by observations, for example, of six men being lifted by the rope of a swinging bell that had been rung by only one of them. However, without paying attention to the interesting question of how this enhancement could be explained, Galileo proceeded directly to the conclusion he was interested in: that the period of the oscillation was completely defined by the setup of the oscillatory system and could not be modified by any external influence. In particular he denied the possibility of modifying the period of a hand-held pendulum (Galilei 1638 p. 98, 1914 p. 97):

> *"First of all one must observe that each pendulum has its own time of vibration so definite and determinate that it is not possible to make it move with any other period than that which nature has given it. For let anyone grasp the cord to which the weight is attached and try, as much as he pleases, to increase or diminish the frequency of its vibrations; it will be time wasted."*

This observation is, in one aspect, not totally wrong. As was described here in paragraph 2.5 above, a pendulum indeed shows oscillations with its natural frequency if it is driven by short pushes with long pauses in-between. However, it was also mentioned that this forced oscillation is not strictly periodic with the natural frequency of the pendulum but with the repetition frequency of the pushes. Moreover, the other assertion of Galileo's is simply false, as was demonstrated by Thomas Young in 1807, almost 200 years after Galileo. Young showed how to make a given pendulum swing with any desired frequency. One only has to apply the periodic driving force sinusoidally in time instead of giving short pushes (see paragraph 6). It would be interesting to ask, but must be left open here, whether the false assertion by Galileo was in some sense the reason why the discovery by Young came so late.

When Galileo drew his conclusion, he had a specific goal: he used it to reject the hypothesis that tides were driven through some mysterious remote action unwanted by him, namely a periodic influence by the moon. His wrong conclusion allowed him to argue that the only cause of the tides is to be found in the motion of the earth (Galilei 1632 p. 426). Thus, the first attempt to relate the tides to the phenomenon of forced oscillations failed; Galileo's theory of tides, seen by him as the desired proof that the earth is moving, was incorrect (Cartwright 2000).

## 4  No resonance with Newton

In light of the concepts of force prevailing at the time of Galileo (Westfall 1971), the phenomenon of forced mechanical oscillations and resonance remained unexplained in his days. Only the concept of *force impressed*, introduced half a century later by Isaac Newton (1643–1727) in his *Principia Mathematica* from 1687 (Newton 1729), is suited for accelerated motion and therefore became the cornerstone of classical mechanics. One could have expected Newton to have tested and demonstrated the power of his new concept in a number of interesting cases, in particular the well-known case of a force acting periodically. Newton, however, did not do so. He did not even treat the free harmonic oscillation of a mass attached to a spring, which recently had been studied successfully by his colleague and competitor Robert Hooke (Hooke 1678). Certainly it is correct to say that Newton in some way covered the harmonic oscillation (Truesdell 1960) when he drew, in a somewhat winding manner, the conclusion from the laws of uniform circular motion that, under an attractive force varying in proportion to the distance (like the elastic force, not mentioned by Newton here), every circular or elliptical motion, and therefore also the limiting case of linear oscillation, takes the same time, thus exactly fulfilling the unique characteristics of the harmonic oscillation. But Newton avoided to mention here any connection to the oscillations of a mass on a spring, and apparently never took a closer look at the way these oscillations, or those of a pendulum, were excited other than by a single push. However, it must be noted that none of the other scientists who investigated oscillations at that time apparently cared about how these oscillations had achieved their amplitude.

On the other hand, Newton certainly was familiar with forces varying periodically in time, namely, with the above-mentioned tidal forces. Among his numerous correct explanations of mechanical phenomena, Newton found that tidal forces are generated in an extended mechanical system when it is rotating in an inhomogenous gravitational field. He treated them not only as causing the tidal deformations of the oceans, along axes oriented with respect to the sun and the moon respectively, but also causing the known periodic perturbations of the moon's motion around the earth.

## 5   Euler's odd motion

One hundred years after Galileo, mechanically forced oscillation and resonance made their next appearance, although not under this name. Around 1739 the resonant excitation of oscillations was rediscovered through the particular phenomenon that two watches firmly fixed to the same massive foundation influence each other when their pendulums have the same length (Krafft 1747; Ellicott 1739). The authors clearly expressed their ignorance of how this could be explained. Leonhard Euler (1707–1783), mathematician at the Petersburg Academy of Sciences, took up this observation when he was working on tides. He realized that the tides were not caused by the vertical component of the periodic tidal force as considered by Newton but by its horizontal component which drives periodic horizontal currents. Attempting to simplify this fairly complicated hydromechanical system, Euler was the first to deal with a driving force varying sinusoidally in time and to write down the differential equation of the (undamped) simple harmonic oscillator under harmonic excitation (Euler et al. 1997). Here it is in a slightly modernized form[1]:

(eq. 7) $$2a\frac{d^2s}{dt^2} + \frac{1}{b}s + \frac{a}{g}\sin\frac{t}{a} = 0$$

This is equivalent to eq. 1 with the friction parameter $c$ set to zero.

Euler's treatment marks a breakthrough in mechanics as the position $s$ of the moving body is expressed for the first time as a function $s(t)$ to be determined by solving the governing differential equation (Katz 1987). Euler also found the general solution to this equation, a remarkable success in those early days of calculus, and he published it under the title *De novo genere oscillationum* (*On a new kind of oscillations*), putting his main emphasis on mathematical details (Euler 1750). His results were totally in agreement with the description given above (for the frictionless case), but were presented in a way that for a modern reader might show a peculiar incongruity between mathematical and physical insight.

In his publication Euler stated (here reported in modern terms) that for parameters $a \neq 2b$ the function $s(t)$ is given by the sum of two harmonic oscillations, $\sin\frac{t}{a}$ and $\sin(\frac{t}{\sqrt{2ab}})$. The first harmonic has its amplitude fixed to $\left|\frac{a^2 b}{g} \frac{1}{2b-a}\right|$ and its phase shift fixed to 0° or 180°, depending on whether $a <2b$ or $a >2b$, while the second harmonic has amplitude and phase shift as freely adjustable parameters. In the special case $a = 2b$, where the two frequencies become equal, Euler found, again correctly, a result equivalent to eq. 5 given above.

---

[1] Euler wrote: $2a\,dds + \frac{s\,dt^2}{b} + \frac{a\,dt^2}{g}\sin.A\frac{t}{a} = 0$.

Thus, Euler discovered that the motion of the forced oscillator in general is given by two harmonics of different frequencies, with the first one equal to the frequency of the external force and the second one determined by the fixed parameters of the system. He further stated that amplitude and phase are fixed in case of the first harmonic but depend on the particular initial conditions for the second.

Thus, Euler had successfully solved the mathematical problem of harmonically driven forced oscillations for the case without friction. He certainly knew the description given by Galileo a century before, and he could have commented on the error it contained, but he didn't. He didn't even comment on the fact that the frequency of the second harmonic (that which may or may not be present in the motion), is the proper frequency of the freely oscillating system, while the other frequency (that of the inevitable oscillation) is the frequency of the driving force. It seems, moreover, that Euler also missed the practical importance of his achievement although he considered the manner to excite motions by periodic forces as an everyday phenomenon and mentioned, among others, tides as an example. He defined these oscillations as that *new kind of oscillations* referred to by the title of his paper but saw his results as a mere curiosity of mathematics:

> *"According to the various relations between the letters a and b, which determine both the acting forces [i.e. the elastic force and the external disturbance - JB], such diverse and astonishing motions were derived which nobody could have expected until the*

> *calculation was completed. With respect to this motion something noteworthy is observed, namely that in a single case the deflections of the body […] steadily increase while the oscillations maintain their constant period. In all other cases, however, the deflections remain finite and of fixed magnitude."* (Letter to Johann Bernoulli, quoted from (Euler et al. 1997, p. 295/303) [2], transl. JB)

In his publication, Euler even added a second proof in order to make his result, which allegedly *nobody could have expected,* more convincing. What he did not mention, however, is the fact that the *fixed* and *finite* amplitude of the forced deflections showing the excitation frequency can take values beyond all limits by simply letting the parameters $a$ and $2b$ approach each other - the essence of every form of resonance in the modern sense of the word. It seems that in his voluminous work which paved the way for the further development of classical mechanics Euler never came back to his *curious result*. Hence, this paper doesn't seem to have attracted any attention at all (and in print hasn't even been translated from Latin yet).

## 6   Laplace's correction

Some 80 years after Euler, Pierre de Laplace (1749–1827) achieved the next remarkable progress in the theory of forced oscillation. Like Euler, he investigated the tides and tried the harmonically driven oscillator as a mechanical model therefor. Laplace did not have the mathematical tools at hand to solve the complicated differential equations for the shape of the ocean surface. So he only gave some general conclusions concerning bodies influenced by a periodic force and additionally by friction or "resistance". Among the basic rules to be applied he stated explicitly *"as a general principle of dynamics"* that

> *"… the state of a system of bodies in which the primitive [i.e. initial – JB] conditions of the motion have disappeared by the resistance [i.e. damping – JB] it suffers, is periodic, like the forces which act on it."* (Cartwright 2000; Laplace 1799, vol I, p. 641)

---

[2] „Prodierunt autem pro varia relatione litterarum $a$ et $b$, a quibus ambae vires sollicitantes pendent, tam diversi ac mirabiles motus, ut eorum indoles nisi calculo peracto praevidiri omnino nequeat. Circa hunc motum id notatu dignum accidit, unico casu spatia per quae corpus C in recta *AB* excurrit perpetuo crescere, oscillationes tamen ejusdem durationis manere: reliquis autem casibus omnibus excursiones esse finitae ac definitae magnitudinis."

He gave no references for this assertion (and no precursor could be found during the present study), but the reason to expose it so clearly was perhaps that the aforementioned misinterpretation of forced oscillations by Galileo was still alive.

Laplace furthermore discovered that the mutual perturbation of the motions of two moons or two planets is gradually increasing over time when their periods are in a proportion of small natural numbers (Laplace 1799). He himself, however, did not make any connection to the phenomenon of resonance. The term *Laplace resonance,* common by now, dates from the mid-20th century only.

## 7   The first full treatment by Thomas Young

When Thomas Young (1773-1829), already famous for his demonstration in 1802 that light is a wave phenomenon, started working on tides, he built his theory explicitly on the basic model of a periodically perturbed pendulum. He was the first to give a complete theory therefor and he also coined the expression *forced vibration* (Young 1807). In a simplified version (Young 1813) he presented a purely physical analysis yielding a very simple explanation of the motion which, as we saw above, to Galileo had appeared impossible, and to Euler odd and unforeseeable.

To put it briefly, Young's analysis starts with a mathematical pendulum of length $L$ and natural frequency $\Omega = \sqrt{g/L}$. ($g$ is the gravitational acceleration.) A point is marked on the pendulum string at an arbitrary height $l$ above the bob, and the motion of the lower part of the pendulum (from that point downward to the bob) is considered separately. The essential observation by Young is that the bob moves exactly as it would if the pendulum had the length $l$ and its suspension point were the moving reference point marked on the string. This motion of the fictitious suspension point of the pendulum of length $l$ is equivalent to a harmonic driving force of frequency $\Omega$ applied to that pendulum. Although the natural frequency of the shortened pendulum is $\omega_0 = \sqrt{g/l}$, this driving force makes it swing with the lower frequency $\Omega$ of the long pendulum. A modern everyday example would be the gentle swinging of a shopping bag in your hand, while holding your arm and your bag in a straight line. (The bag alone represents length $l$, arm plus bag the length $L$.) This example was addressed almost literally by Young himself, in order to emphasize how easy it is to modify the frequency of a handheld pendulum. Young did not make explicit that hereby he was disproving Galileo.

Young extended his mechanical analogy easily to the case $l > L$. He only imagined the string of length $L$ extended beyond its real suspension point, allowing him to mark an imagined reference point in this upper section. This point also moves harmonically, but in opposite phase (remember the "indirect oscillation"), and can be taken as the suspension point of a fictitious pendulum of longer length $l > L$.[3] In both cases the ratio of the amplitudes of the oscillations of the bob and its moving fictitious suspension point are related by simple geometrical reasoning to the ratio of the two lengths involved, $L$ and $l$, which in turn is related to the ratio of the two frequencies $\Omega$ and $\omega_0$. This results in $L/l = \omega_0^2/\Omega^2$, from which Young immediately got the correct formulae for amplitude and phase of the stationary forced oscillation (in the frictionless case). Since Young was only interested in tides, he confined his studies to the stationary state of the oscillation and consequently missed the exceptional case of resonance of the frictionless oscillator where no stationary oscillation exists.

---

[3] An animated illustration is seen at https://de.wikipedia.org/wiki/Erzwungene_Schwingung (visited 4/11 2017)

Young subsequently augmented his analysis by considering *resistance* (i.e. friction), as he explicitly addressed it in the title of his publication (Young 1813). Herein he also discussed the phase shift and the transition to the stationary oscillation. Finally, in a supplement to his article *Tides* in the Encyclopedia Britannica of 1824, he gave the analytical solution to the governing differential equation (eq. 1), equivalent to eqs. 2 to 4 above and differing only in notation (Young 1824).

With respect to understanding the tides, the work of Young marked significant progress, although his equations, based on the damped simple pendulum, were far too simple for the calculations of high and low water to be useful in practice. Hence, Young's approach was the last attempt to interpret the tidal motion in terms of a one-dimensional oscillator. Less than 20 years later a detailed 3-dimensional treatment was published by G. B. Airy in the London Encyclopedia Metropolitana under the title *Tides and Waves*. Young's work was not even quoted there (Airy 1841).

With respect to the development of classical mechanics, Young was treated even worse. His breakthrough in the problem of forced oscillations remained completely unnoticed. This seems remarkable, given that his article *Tides* with the first full solution to the problem of forced oscillations with damping, was reprinted in the Encyclopedia Britannica (under the keyword *Thomas Young*) for about 50 years. It seems that Young's mathematical result never has been quoted at all, whereas his earlier mechanical explication of the frictionless case received, at least, one quotation: a footnote to the article on elastic oscillations in the Encyklopädie der Mathematischen Wissenschaften by Horace Lamb (Klein and Müller 1901, vol IV.4, p. 225, FN 19).

This story resembles the neglect of the *new kind of oscillations* published by Euler some 80 years earlier. But forced oscillations had to wait another 80 years after Young's publication to attract broader recognition, as indicated e. g. by their appearance in textbooks on mechanics. Nothing is found on this subject in the famous textbooks by Lagrange, Laplace, Francoeur, Poisson, and Poncelet from the then leading French school of mechanics (Lagrange 1788; Laplace 1799; Francœur 1804; Poisson 1811; Poncelet and Kuppler 1840). The same must be said for the later textbooks in German by Kirchhoff, Mach, and Hertz (Kirchhoff 1876; Mach 1883; Hertz 1894). Also, the 1400-pages *History of Physics* by Ferdinand Rosenberger, published from 1882 on, ignores the phenomenon (Rosenberger 1882). In general, the term *resonance* remained reserved for the acoustic phenomenon and was used in other areas in a figurative sense only. In optical spectroscopy, e. g., the term resonance had been suggested by G. G. Stokes in order to "*illustrate*" that spectral lines of absorption and emission have the same wavelength, thus the same frequency (Stokes 1860).

## 8  Acoustic resonance: Seebeck, Helmholtz, and Rayleigh

Meanwhile, studies of acoustical resonance had revealed that a basic role is played here by simple mechanical resonance. This was established by August Seebeck (1805-1849), son of the discoverer of the thermoelectric effect and at this time president of the institute which later became the Technische Universität Dresden. Seebeck presented a calculation of the resonant response of a rigid plate elastically fixed to a position of rest and subjected to an acoustical plane wave (Seebeck 1844). Apparently again independent of both Euler and Young, he derived the complete solution for the one-dimensional forced oscillation. In passing he also gave the formula valid for an external force with arbitrary time dependence, using a general mathematical procedure first proposed in 1830 by Jean-Marie Duhamel in the context of heat conduction (Duhamel 1834, 1833) and not employed again until 1883, this time in the theory of elasticity (Clebsch et al. 1883, p. 538). Seebeck's publication again passed by totally unnoticed. It has since been quoted only once, namely, in a footnote of the influential 20$^{th}$ century textbook on acoustics by Ferdinand Trendelenburg (Trendelenburg 1961). Obviously unaware of the work by Young, Trendelenburg states incorrectly that Seebeck was the creator of the theory of forced oscillations, while he deplores correctly that Seebeck's important paper had fallen into oblivion.[4]

Only twenty years later, however, time had come for mechanical resonance to make its appearance. In his exhaustive publication on acoustics, Hermann von Helmholtz (1821–1894), already famous for finding the general law of energy conservation, treated acoustic resonance in terms of the basic model of simple forced oscillation (Helmholtz 1863 p. 60 ff). Here, Helmholtz presented an explanation of the whole phenomenon of mechanical resonance in simple words, well suited for the broader public, but also gave the full mathematical treatment like our eqs. 1 to 5 in an appendix. However, no precursor was mentioned besides a global reference to what "*theoretical mechanics*" would tell (Helmholtz 1863 p. 64).

---

[4] "Die Theorie der erzwungenen Schwingung wurde von A. Seebeck […] ausgearbeitet. Diese, eine Fülle wichtigster Erkenntnisse enthaltende Arbeit scheint völlig in Vergessenheit geraten zu sein, man findet sie nirgends zitiert."

Helmholtz also emphasized that the phase shift $\delta$ is of central importance for the amplitude of the forced oscillation. He further pointed out that the low frequency beats, which are not difficult to hear at the beginning of excitation of a tuning fork if the exciting frequency is close to resonance, are caused by superposition of the stationary forced oscillation with a damped proper oscillation.

The next one to elaborate on simple mechanical forced oscillations was Lord Rayleigh in his two volumes *Theory of Sound* (Rayleigh 1878). He exposed the formulae giving reference to Helmholtz, but again not to any other earlier work. It is quite ironical that the same Rayleigh wrote about Thomas Young in another place, with respect to his earlier geophysical investigations (Rayleigh 1964 p. 197):

> *The case of Young will at once suggest itself as that of a man who from various causes did not succeed in gaining due attention from his contemporaries. [Scientific - JB] Positions which he had already occupied were in more than one instance reconquered by his successors at a great expense of intellectual energy.*

It is noted that Rayleigh also pointed out that the fundamental role of mechanical resonance reaches far outside the field of acoustics. As an example he mentioned the rolling of a ship in heavy sea.

## 9   Electromagnetic resonance: Thomson, Oberbeck, Hertz

At this time, the significance of resonance was also noted – albeit still not under this name – in connection with electromagnetic processes. In 1853 William Thomson, later famously known as Lord Kelvin (1824–1907), one of the leading physicists of the entire 19th century, analyzed the discharge of a Leyden jar through a simple wire and found the equations for what today is well known as the LC circuit (Thomson 1853). The equation for its resonance frequency ($\omega = \sqrt{LC}$ ) in some books still bears the name *Thomson formula*.

The first author to cautiously assign the word "resonance" to this electrical phenomenon was Anton Oberbeck (1846–1900), then professor of physics at the Universität Greifswald. In his 1885 paper *"On a phenomenon with electrical oscillations which is similar to resonance" [transl. JB]* he derived the equations corresponding to our eqs. 1 to 4 above and stated that one finds here the same formulae and phenomena as with any other type of resonance (Oberbeck 1885). He also was the first to record the resonance curve, in the form of data tables of the voltage excited at different frequencies. It was this resonant enhancement of an alternating voltage induced in an LC circuit which achieved unforeseen importance when it was employed in 1886 by Heinrich Hertz (1857–1894) to generate and detect free electromagnetic waves (Hertz 1887)(Hertz 1888). These, in turn, were shown to enable wireless communication by Guglielmo Marconi from 1895 on. The 1887 publication by Hertz shows the first graph of the resonance curve (see Fig. 2) that Richard Feynman some 75 years later would see in every volume of *The Physical Review*.

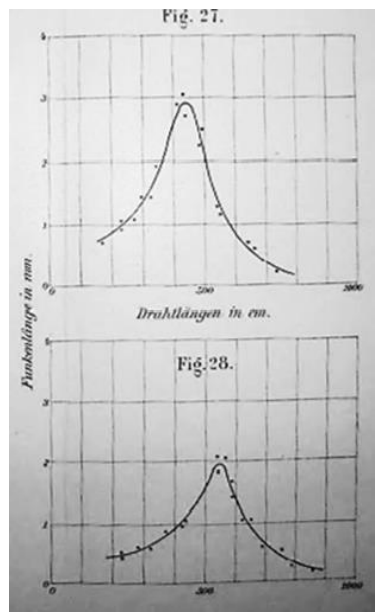

Fig.2 The first two graphs of the resonance curve published (Hertz 1887)

## 10  Applied mechanics: Redtenbacher, Radinger, Sommerfeld, and others

In the field of applied mechanics, concerned mainly with construction and mechanical engineering, the design of new edifices and machines had for centuries been based solely on considerations of the static equilibrium of forces. In particular, little attention was paid during the $19^{th}$ century to theoretical or mathematical mechanics (Lorenz 1902). An example is once again the reluctant reception of forced oscillations and mechanical resonance. Certainly there is impressive evidence of the state of the art reached by using statics alone, like the steel constructions of the Brooklyn Bridge from 1883 and the Eiffel Tower from 1889. Apparently, the obligatory safety margins were sufficient to cope with the excess loads of dynamical origin so that no need was felt to transcend the purely static calculation. This opinion had even survived the experts' discussions triggered by the breakdown of suspension bridges, e. g. the disaster of the Broughton bridge in England back in 1831, which obviously was caused by vibrations provoked by a cohort of soldiers enjoying the resonance effect of marching in lockstep (Phil. Mag. editorial 1831).

Consequently, in the *History of applied mechanics* of 1885 by Rühlmann forced oscillations are not even mentioned; and the overall subject *oscillations* appears there for the last time in the context of the pendulum experiments made by Huygens in the $17^{th}$ century (Rühlmann 1885). It is noted that, even today, the rare textbooks on applied mechanics that contain some information about the historical development don't comment on this history (e.g. Szabo 2013a, 2013b).

Observations of unwanted oscillations were frequently made during the 19th century, mainly with the ever faster running fixed and mobile steam engines. Their oscillating pistons and piston rods inevitably generate a periodic disturbance of the rest of the machine. In 1855 such oscillations had already been identified and analyzed as a type of forced oscillation by Redtenbacher (1809–1863), then professor of mechanics at the precursor of today's Karlsruhe Institute of Technology and author of the standard text book on steam locomotives (Redtenbacher 1855). He repeated, seemingly independently of the works mentioned above, the calculations performed first by Euler, i.e. for undamped systems. Like Euler more than 100 years earlier, Redtenbacher was surprised by the mathematical result that at the exact resonance frequency the amplitude is steadily increasing with time. Like Euler, he felt the need to confirm this through an independent calculation using a different method. Redtenbacher surpassed Euler, however, when he gave a clear description of what happens in practice when the resonance frequency is approached and finally passed. In his words, the oscillations of the steam engine become alarming when

> *"the speed of the engine approaches a value that makes the period of revolution of the drive axle come close to that of a proper oscillation. Therefore, every engine has a certain rotational speed which provokes a strong vibration of the construction. If a locomotive gradually changes its motion from slow to extraordinarily fast, the vibration will be weak at the beginning, then growing stronger and stronger; however, after having luckily passed this dangerous instant, the vibration will diminish again while the speed continues to increase, to finally disappear completely at infinite speed so that the engine would stand completely quiet."* (Redtenbacher 1855 p. 150, transl. JB)

Redtenbacher correctly drew the connection to the resonant excitation of oscillations in pendulums and steamboats, and he even mentioned that this phenomenon bears some resemblance to what Laplace had discovered with the mutual disturbance of the planets' orbits. However, his analysis was rejected by others (Zech 1867) and apparently had no further consequences. His critics argued that the total transfer of momentum in each period would sum up to zero. This can be true (see chapter 2.4 above), but only in the frictionless case and never at the exact resonance frequency.

As late as 1900 two outstanding mathematicians from Göttingen, Felix Klein (1849–1925) und Arnold Sommerfeld (1868–1951), who were actively trying to bring mathematics, physics and engineering into closer connection, had to state that applied mechanics unfortunately had been dominated by the notion of static equilibrium of forces (Klein 1900). As Sommerfeld expressed it pointedly, it was Johann Radinger (1842–1901), one of the leading machine designers in those days,

> *"…who must be credited for having awoken the engineers' dynamic conscience. He rediscovered in mechanical engineering the basic law by Newton that mass times acceleration equals force."* (Sommerfeld 1903 p. 609[5], transl. JB)

According to the textbook *High speed steam engines* by Radinger, such *mass forces* as generated by any accelerated motion may cause mechanical stress in levers, piston rods, transmission belts or drive shafts that is several times stronger than if calculated with static loads only (Radinger 1892).

Even then, however, the problems special to forced oscillations were still overlooked. The engineering calculations were based on the general assumption that the energy transmitted in one quarter of an oscillation would be given back in the next quarter (Weisbach 1850; Radinger 1892). This can be true, but only for phase shift $\delta = 0°$ or $180°$. This in turn only holds, if a frictionless oscillator is in stationary forced oscillation driven at a frequency different from its proper frequency. It is definitely false for every oscillator at resonance, with or without friction, as had been pointed out already by Young in 1823 and by Helmholtz in 1863. In this case there is no energy flux at all from the oscillator back to the source of the driving force. Since the early treatments of forced oscillations frequently were based on just these simplifying assumptions, it seems possible that the role of the phase shift between driving force and stationary oscillation had been overlooked.

---

[5] „Radinger gebührt das große Verdienst, das dynamische Gewissen des Technikers geweckt zu haben. Er entdeckte im Maschinenbau den Newtonschen Grundsatz von neuem, wonach Masse x Beschleunigung = Kraft ist."

Outside the exact resonance, however, the energy flows to and fro, and when friction is included, the net energy transfer to the oscillator during stationary oscillation is always positive since the phase shift always lies somewhere inside the interval from 0° to 180° (see eq. 6). This basic insight that the energy transfer makes the stationary oscillation stationary because it balances the energy loss due to friction was expressed in a general textbook on mechanics for the first time in 1902 (Riecke 1902 p.95).

Unexpected vibrations and even breakdown and destruction continued to occur with steam engines and in long drive shafts in high speed operation like in steamboats, as well as in railway bridges, like in a particularly catastrophic way in 1891 in Münchenstein (Switzerland). Shortly after 1900, renowned mechanical engineers like August Föppl (professor at the Technische Universität München, see Eckert 2013b, p. 124) and Hermann Frahm, a leading shipbuilding engineer, began to look for an explanation based on resonance. Once again, their wording indicates that an apparently new concept was proposed for consideration:

*"The only option is to consider a force dynamic in nature as the main cause of the destructions […] It has already been suggested from various sides that the possibility of so-called resonance oscillations in the drive shafts needs to be considered, which, however, at first were only speculations without base."* (Frahm 1902 p. 797[6], transl. JB)

Frahm also turned the question around and applied resonance to reduce forced oscillations. He noticed that a resonantly oscillating water reservoir onboard a resonantly rolling ocean liner would oscillate 180° behind the swell, thus reducing the net driving force considerably. His invention of antiroll tanks revolutionized the construction of ocean liners and even stimulated the nomination of Frahm for the 1913 Nobel prize in physics, put forward by no lesser scientist than Svante Arrhenius (Nobel committee 1913).

## 11 The "Sommerfeld-Effect"

Arnold Sommerfeld, after being appointed the chair of applied mechanics at the *Technische Hochschule Aachen* (the first appointment of a physicist and mathematician on a chair of such technological significance), strongly urged the reluctant engineers to finally accept the significance of mechanical resonance. He even demonstrated it in an experiment (Sommerfeld 1902), today sometimes known as *Sommerfeld effect* (Eckert 1996, 2013b, 2013a). It consisted of a wobbly table that supported a heavy machine running at increasing rotation speed due to increasing power input. However, shortly before the resonance frequency of the

---

[6] "Es blieb nur übrig, bisher nicht genügend erkannte Kraftwirkungen dynamischer Natur als die hauptsächlichste Ursache jener Zerstörungen anzunehmen, in Verbindung natürlich mit schädlichen Einflüssen anderer Art, wie Korrosionen u. dergl.
Von verschiedenen Seiten war schon auf die Möglichkeit des Auftretens von sogenannten Resonanzschwingungen in den Wellenleitungen hingewiesen worden, doch waren dies nur Mutmaßungen, die zunächst noch jeder Unterlage entbehrten."

table with the machine on it was reached, the rotation speed ceased to increase, and further growth of energy input only served to increase the amplitude of the unwanted oscillation. Sommerfeld did not fail to say that this would mean an increase of the fuel bill without getting anything but the risk of damaging the machine and the building. Sommerfeld later moved to the University of Munich and founded a school of theoretical physics which was to become one of the most influential ones for the development of quantum mechanics.

Klein managed to establish a new chair for *Technische Physik* at the University of Göttingen and secured the appointment of Hans Lorenz (1865–1940), a mechanical engineer with an exceptionally solid competence in mathematics. Lorenz wrote the textbook *Engineering Mechanics of technical systems* (Lorenz 1902) where in the preface he credited himself for extensively treating "*the forced and damped oscillations having only recently gained such an importance*" (transl. JB). He also presented in detail the controversial history of the complicated relations between mathematical mechanics and applied mechanics during the entire 19$^{th}$ century. It fits into this picture that the first full mathematical treatment of the "Sommerfeld effect", published in 1903 by Radakovic (Radakovic 1903) who used advanced techniques of analytical mechanics, passed by as unnoticed as his similar treatment of the vibrations of railway engines (Radakovic 1906).

Klein, Sommerfeld and others from 1901 on organized the *Encyklopädie der mathematischen Wissenschaften mit Einschluss ihrer Anwendungen (Klein and Müller 1901)*, precursor of the widely known *Handbook of Physics*. Here the forced mechanical oscillations of a point mass and the resonance are presented in appropriate depth (Karman 1910; Mises 1911). The first dedicated textbook seems to be the small book *Technische Schwingungslehre* by Wilhelm Hort (Hort 1910). It presents mechanical and electrical oscillations on the same basis and shows, in the 2$^{nd}$ edition from 1922 (Hort 1922), the by now well-known two sets of curves for amplitude and phase shift of the stationary forced oscillation as functions of the driving frequency for various degrees of damping (eqs. 3 and 4).

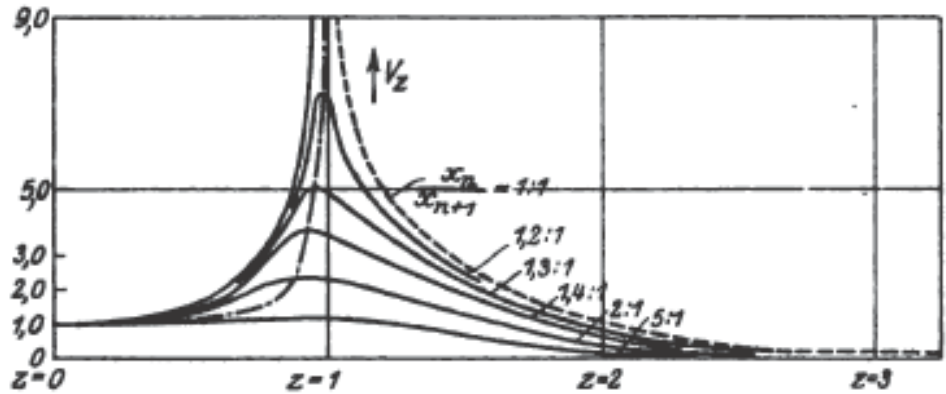

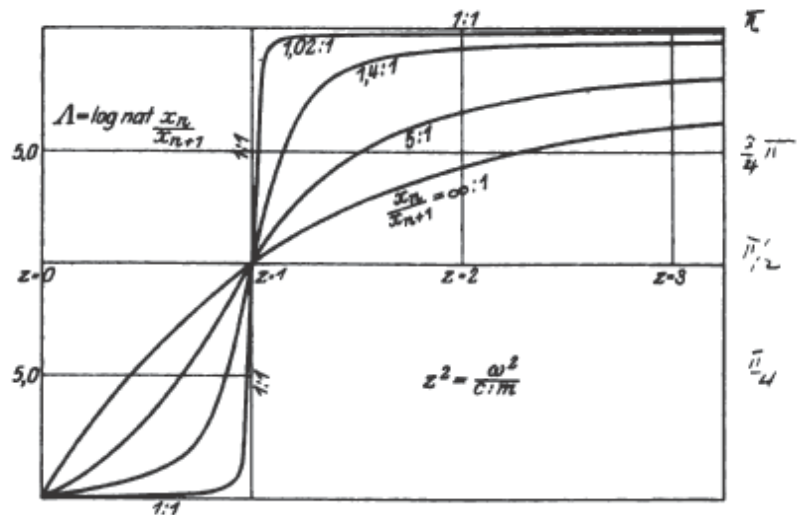

Fig. 3 The first complete set of graphs for amplitude and phase of the stationary forced oscillation (Hort 1922)

An exhaustive standard textbook, *Vibration Problems in Engineering,* was published in 1928 by Stephen Timoshenko (Timoshenko 1928). Thus, forced oscillations and mechanical resonance finally had come into the engineer's field of view.

However, one associated problem was still overlooked, since the treatment was always limited to the stationary state of the forced oscillation. None of the mentioned textbooks pointed out the aggravating fact that during the transient state, even if the oscillator starts from zero displacement and velocity, its amplitude can grow temporarily to almost twice the amplitude of the final stationary oscillation. This was not only a clear consequence of the formulae but had also been demonstrated in experiments on tuning forks (Hartmann-Kempf 1904). Timoshenko later generalized the model of harmonic oscillations forced by a harmonic driving force to the advanced picture of *dynamical response* of an elastic structure to a force varying arbitrarily in time. With this method, an appropriate treatment of the transient state is automatically included. We note in passing that this stage of development had already been reached back in 1844 when Seebeck used the same mathematical technique today called Duhamel's integral.

Motivated by various earthquake disasters between 1900 and 1933, this advanced concept was then tried in analyzing the stability of constructions subjected to variable forces (Herzog 2009). An appropriate treatment of the dynamical response of buildings, however, was achieved in civil engineering only in the second half of the $20^{th}$ century, when the licensing procedure for nuclear power plants required that earthquake safety was proven. Meanwhile, several more disasters had occurred that were associated with the underestimation of dynamical load caused by earthquake (Housner 2002).

## 12  Late response in Physics too

In physics, too, the general interest in the resonance phenomenon grew slowly. It was only in 1896 that Max Wien (1866-1938), in later years professor of physics in Danzig and Jena, discovered the interesting fact that there had been a confusion between two slightly different resonance frequencies: The stationary forced oscillation has maximum energy for driving frequency $\Omega = \omega_0$ but maximum amplitude for driving frequency $\Omega = \sqrt{\omega_0^2 - 2\beta^2}$, which is even lower than the frequency of the free damped oscillation, $\omega_d = \sqrt{\omega_0^2 - \beta^2}$ (Wien 1896). The lack of interest among the physicists is also seen in the main reference textbook of physics of the time, the multivolume 5th edition of "*the Müller-Pouillet*" of 1905. It mentions resonance in quotation marks and only in the chapter on acoustics (Müller et al. 1905, p. 601). Forced mechanical oscillations were neglected until the next edition, published in 1929. The same holds for other German textbooks, e.g. for the widely used *Lehrbuch der Physik* by Grimsehl.

Then, however, the situation changed rapidly. Only a couple of years later, the importance of mechanical resonance as a model process in various fields of physics and engineering was reflected even in popular science books for the broad public (e.g., Karlson 1934). It can be supposed that, besides the increasing use of wireless broadcasting, the frequent use of the term resonance in quantum mechanics also might have contributed to this rapid change (see e.g. Heisenberg 1926). Here, the relation $E = \hbar\omega$, first encountered by Max Planck in 1900 (Planck 1900), establishes a universal correspondence of frequency and energy which makes the concept of two matching frequencies applicable to the case of two matching energy levels. Although the quantum mechanical case differs from the mechanical one insofar as the oscillating quantities are not position and momentum of a mass point but the real and imaginary components of the wave function, which are not governed by eq. 1 but by the Schrödinger equation, their amplitude and phase are given by the same formulae eq. 3 and 4 as derived for the forced mechanical oscillation. An example is seen in the two resonance curves which Feynman had used in 1962 to emphasize the fundamental importance of this common concept (see Fig. 1). Both curves show the resonance at 1520 MeV between the rest energy of an unstable particle and the total center-of-mass energy of a certain pair of two other particles which in collision can combine to form it. The amplitude of the forced oscillation excited in such collision indicates the probability that the new particle is formed, which in turn is measured through the reaction cross sections as shown in the figure. The particle was baptized with the name "Λ(1520) resonance".

## 13  Conclusion

A review of literature on resonance phenomena shows that the simple one-dimensional mechanical forced oscillation, which already was discovered in the early 17$^{th}$ century, was fully recognized only in the middle of the 20$^{th}$ century as a fundamental model for a broad variety of processes in physics and engineering.

Acknowledgement

Helpful discussions with Falk Riess and Rainer Bleck are gratefully acknowledged. The author, being a physicist not specialized in historical sciences, would appreciate receiving suggestions of peer reviewed journals possibly inclined to publish this review.